\documentclass[aps,prl,twocolumn,groupedaddress,showpacs]{revtex4}

\usepackage{graphicx}
\usepackage{dcolumn}
\usepackage{bm}

\begin{document}

\title{Magnetic field control of elastic scattering in a cold gas of fermionic
lithium atoms}

\author{S. Jochim}
\author{M. Bartenstein}
\author{G. Hendl}
\author{J. Hecker Denschlag}
\author{R. Grimm}

\affiliation{Institut f\"ur Experimentalphysik, Universit\"at
Innsbruck, Technikerstr.\ 25, 6020 Innsbruck, Austria}

\author{A. Mosk}
\affiliation{FOM instituut voor plasmafysica Rijnhuizen, P.O.\ Box
1207, 3430 BE Nieuwegein, The Netherlands}

\author{M. Weidem\"uller}
\affiliation{Max-Planck-Institut f\"ur Kernphysik, Postfach
103980, 69029 Heidelberg, Germany}

\date{July 24, 2002}

\begin{abstract}
We study elastic collisions in an optically trapped spin mixture
of fermionic lithium atoms in the presence of magnetic fields up
to 1.5\,kG by measuring evaporative loss. Our experiments confirm
the expected magnetic tunability of the scattering length by
showing the main features of elastic scattering according to
recent calculations. We measure the zero crossing of the
scattering length that is associated with a predicted Feshbach
resonance at 530(3)\,G. Beyond the resonance we observe the
expected large cross section in the triplet scattering regime.
\end{abstract}

\pacs{34.50.-s, 05.30.Fk, 39.25.+k, 32.80.Pj}

\maketitle

In an ultracold atomic gas, the $s$-wave scattering length $a$
characterizes the elastic interactions and is therefore an
especially interesting parameter to manipulate. Changes of $a$
have a profound effect on the dynamics of the gas and on its
thermodynamic stability. For spin mixtures of fermionic atoms, the
scattering length governs the stability and critical temperature
of a Cooper-paired superfluid phase. The scattering length can be
conveniently tuned if it depends on the magnetic field, as is the
case near a Feshbach resonance. For bosonic atoms, such resonances
have been observed \cite{inouye1998a, courteille1998a,
vuletic1999a}, and used for attainment and manipulation of a BEC
in $^{85}$Rb \cite{cornish2000a} and for the production of bright
solitons in bosonic $^{7}$Li \cite{khaykovich2002, strecker2002}.

For fermionic gases, Loftus {\it et al.}\ have recently observed a
Feshbach resonance between two different spin states of $^{40}$K,
which changes scattering properties in a narrow magnetic field
range \cite{loftus2002}. The other fermionic species currently
used in several experiments, $^6$Li, is predicted to have a very
broad Feshbach resonance which completely changes the character of
interactions in a wide experimentally accessible field range.
Magnetic-field tuning of elastic scattering at small fields far
below the resonance was recently used by Granade {\it et al.}\ to
obtain a sufficient scattering cross section  for the all-optical
production of a degenerate Fermi gas of lithium
\cite{granade2002}. At higher magnetic fields, an expected extreme
tunability promises great prospects to observe new phenomena,
e.g.\ in Cooper pairing \cite{Stoof1996a, Holland2001}, cold
molecule formation \cite{donley2002} and studies of Fermi gases in
reduced dimensionality \cite{Petrov2000a}.


In this Letter, we experimentally explore the magnetic tunability
of elastic scattering in a spin-mixture of fermionic lithium atoms
in high magnetic fields up to 1.5\,kG.  Our results verify the
expected dependence of the cross section that follows from
theoretical calculations of the scattering length
\cite{Houbiers1998a, kokkelmans2002, venturi2002}. As a particular
feature associated with a predicted Feshbach resonance
\cite{Houbiers1998a}, we observe the zero crossing of the
scattering length that occurs at a field of 530\,G. The exact
location of this feature is of great interest as a sensitive input
parameter to better constrain the uncertainty in the molecular
potentials for more accurate theoretical calculations of the
scattering properties of $^6$Li. Our measurements of elastic
collisions are based on evaporation out of an optical dipole trap.
We use an optical standing-wave trap developed earlier
\cite{Mosk2001}, which employs the power enhancement inside an
optical resonator to create a deep trap with a sufficiently large
volume to capture a large number of lithium atoms.

\begin{figure}
\includegraphics[width=8cm]{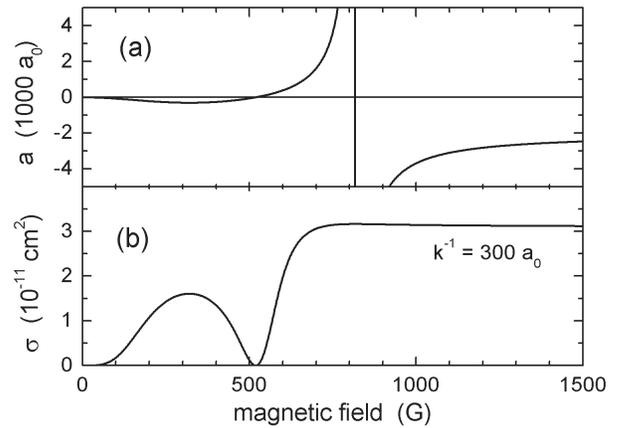}
\caption{\label{resonance} (a) Model curve approximating the
results of \cite{Houbiers1998a, kokkelmans2002, venturi2002} for
the $s$-wave scattering length of $^6$Li atoms in the two lowest
spin states versus magnetic field. (b) Corresponding behavior of
the scattering cross section at a finite collision energy with a
relative wave number of $k = (300 a_0)^{-1}$.}
\end{figure}

The scattering properties in different spin-mixtures of fermionic
lithium atoms were theoretically investigated by Houbiers {\it et
al.}\ \cite{Houbiers1998a}, Kokkelmans {\it et al.}
\cite{kokkelmans2002}, and Venturi and Williams
\cite{venturi2002}. Magnetic tunability, of particular interest
for Cooper pairing in a Fermi gas \cite{Stoof1996a, Holland2001},
was predicted for the stable combination of the two lowest states
$|1\rangle$ and $|2\rangle$; at low magnetic field these states
correspond to $F=1/2$, $m_F=+1/2$ and $m_F=-1/2$, respectively.
Most prominently, a broad Feshbach resonance is expected to mark
the transition from the low-field scattering regime to the
high-field region. In more detail, the $s$-wave scattering length
$a(B)$ is expected to show the following behavior: After being
zero in the absence of a magnetic field, a local maximum of
$|a(B)|$ occurs with rising field at $\sim$300\,G with $a \approx
-300\,a_0$, where $a_0$ is the Bohr radius. Then, as a precursor
of the Feshbach resonance, the scattering length crosses zero in
the range between 500\,G and 550\,G. After passing through the
resonance somewhere between 800\,G and 850\,G scattering in higher
fields is dominated by the triplet potential with a very large and
negative scattering length. The available theoretical data
\cite{Houbiers1998a, kokkelmans2002, venturi2002} show the same
behavior for $a(B)$ within some variations due to the limited
knowledge of the molecular interaction parameters.
Fig.~\ref{resonance}(a) illustrates these predictions for the
scattering length $a(B)$ by a corresponding model curve
\footnote{We use the analytical function
$a(B)=(1-\exp(-(B/p_1)^2))\times(p_2(1+p_3/(B-p_4)-p_5\arctan(B-p_4)))$
with $p_1=279$, $p_2=-1528$, $p_3=202$, $p_4=816$, $p_5=312$ to
approximate the calculated $a(B)$ of Refs.~\cite{Houbiers1998a,
kokkelmans2002,venturi2002}.} that approximates the results of
Refs.~\cite{Houbiers1998a, kokkelmans2002, venturi2002}.

In a cold gas at finite temperature the cross section for elastic
scattering of non-identical particles is unitarity-limited to a
maximum value of $\sigma_{\rm max} = 4\pi/k^2$, where
$k=mv/(2\hbar)$ is the wavenumber corresponding to a relative
velocity $v$ and a reduced mass $m/2$. Taking into account the
B-field dependent scattering length $a(B)$ and the unitarity
limit, the resulting B-field dependent cross section can be
written as $\sigma = 4\pi a^2/(1 + k^2a^2)$. For the considered
$|1\rangle - |2\rangle$ spin mixture of $^6$Li the expected
behavior of the cross section is shown in Fig.~\ref{resonance}(b)
for the example of a wave number $k=(300\,a_0)^{-1}$ close to our
experimental conditions. Most notably, as a consequence of the
unitarity limit in combination with the very large scattering
length for high magnetic fields, the Feshbach resonance does not
appear as a pronounced feature in the cross section.
The zero crossing of the scattering length, however, leads to a
vanishing scattering cross section and thus shows up as a
manifestation of the resonance.

The resonator-enhanced dipole trap (REDT) \cite{Mosk2001} makes
use of the enhancement of the laser intensity inside a linear
optical resonator to create a large and deep trapping volume for
lithium atoms. The power provided by a 2-W Nd:YAG laser (Innolight
Mephisto-2000) at a wavelength of 1064\,nm is enhanced by a factor
of 120 to create a far red-detuned 1D optical lattice trap with an
axial period of 532\,nm and a transverse $1/e$-radius of
115\,$\mu$m. The maximum trap depth is of the order of 1\,mK. To
vary the trap depth the resonator-internal power is
servo-controlled by an acousto-optical modulator in the incident
laser beam. From a standard magneto-optical trap (MOT) operated
with diode lasers we typically transfer $5\times10^5$ $^6$Li atoms
into roughly 1000 individual wells at a temperature of
$\sim$\,400$\mu$K. The resulting peak density is
$\sim$\,$1.5\times10^{11}\,$cm$^{-3}$. By extinguishing the
repumping light of the MOT 1\,ms before the main trapping light is
turned off, all atoms are pumped into the two states $|1\rangle$
and $|2\rangle$ to create the desired spin mixture.

The magnetic field is produced by a pair of water-cooled coils
outside of the glas vacuum cell of the trap. At a maximum
continuous operation current of 200\,A the coils produce a
magnetic field of 1.5\,kG with a curvature of only 75\,G/cm$^2$
along the symmetry axis; the corresponding power dissipation is
6\,kW. The set-up allows for a maximum ramp speed of 5\,G/ms
within the full range. The magnetic field is calibrated by
radio-frequency induced transitions from $|2\rangle$ to the state
that at $B=0$ corresponds to $F=3/2$, $m_F=+1/2$. The latter is
unstable against inelastic collisions with $|2\rangle$ which leads
to easily detectable loss. With a fit to the Breit-Rabi formula we
obtain a calibration of the magnetic field to better than 1\,G
over the full range.

\begin{figure}
\includegraphics[width=8cm]{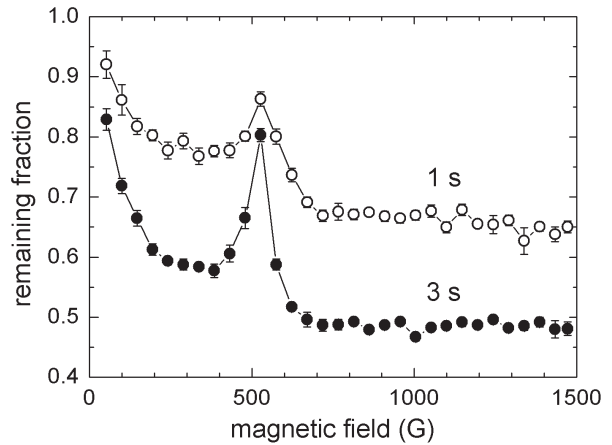}
\caption{\label{plain}Evaporative loss measurements over the full
magnetic field range. The data points show the measured number of
atoms remaining in the trap after 1\,s ($\circ$) and 3\,s
($\bullet$) of plain evaporation at a constant trap depth of
750\,$\mu$K.}
\end{figure}

The basic idea of our measurements is to observe elastic
collisions through
evaporative loss at a
variable magnetic field \cite{Chin2000}. Our trap is initially
loaded under conditions where the effective temperature $T$ of a
truncated Boltzmann distribution \cite{Luiten1996a} is only
slightly below the trap depth $U$. A strongly non-thermal
distribution is thus created with a small truncation parameter
$\eta = U/k_BT \approx 2$. The thermal relaxation resulting from
elastic collisions then leads to rapid evaporative loss and
cooling of the sample, i.e.\ an increase of $\eta$. The trap depth
can be kept constant to study plain evaporation or, alternatively,
ramped down to force the evaporation process.

In a series of plain evaporation experiments performed at a
constant trap depth of 750\,$\mu$K we measure evaporative loss
over the maximum accesible range of magnetic fields up to 1.5\,kG.
After a fixed holding time the remaining atoms are retrapped into
the MOT and their number is measured via the flourescence signal
by a calibrated photodiode. The loss signal is recorded for
holding times of 1\,s and 3\,s corresponding to the time scale of
evaporation. These holding times are short compared with the
rest-gas limited lifetime of 30\,s. Fig.~\ref{plain} shows the
result of 1000 different measurements obtained in an acquisition
time of six hours. The data points are taken in a random sequence
for 31 magnetic field values equally distributed over the full
range. Data points for 1\,s and 3\,s are recorded alternatingly.
This way of data taking ensures that the signal is not influenced
by residual long-term drifts of the experimental conditions.

The observed evaporation loss in Fig.~\ref{plain} shows a
pronounced dependence on the magnetic field which is to be
compared with the expected cross section for elastic collisions as
displayed in Fig.~\ref{resonance}(b). After being very small at
low magnetic fields, the loss increases for fields up to
$\sim$350\,G where an expected local maximum of evaporative loss
is observed. The loss then decreases and disappears at about
530\,G as a consequence of the predicted zero crossing of the
scattering length. Here the slight observed loss in the 1\,s curve
is explained by the finite ramp time of the magnetic field. In the
100\,ms ramping time some evaporation does already take place. At
530\,G the decrease of the trapped atom number between 1\,s and
3\,s is fully explained by rest gas losses without any further
evaporation. For higher magnetic fields evaporative loss rapidly
rises until it levels off at about 700\,G. Up to the maximum
attainable value of 1.5\,kG high evaporation loss is observed. A
slight decrease of the atom number for fields exceeding 1\,kG
occurs which we attribute to technical reasons; we observe an
increasing noise for currents higher than $\sim$130\,A in the
error signal of the resonator lock of the REDT. The relatively
large and constant evaporative loss for fields exceeding 700\,G is
consistent with the predicted behavior of the cross section.

The evaporative cooling effect is confirmed by measuring the
change of the truncation parameter $\eta$ after 3\,s of trapping
at selected values of the magnetic field. For thermometry we turn
off the magnetic field to avoid further elastic collisions and
adiabatically lower the trap depth in a 1-s exponential ramp. The
fraction of remaining atoms as a function of the relative depth
then provides a good measure of $\eta$. At the zero-crossing at
530\,G we observe only a slight increase of $\eta$ to a value of
2.3(2) which is explained by the unavoidable evaporation during
the magnetic field ramps. At $340\,$G close to the local maximum
of $|a|$ we find an increase of $\eta$ to 4.2(3) as a clear
evidence of evaporative cooling. At $720\,$G, i.e.\ in the case of
a large positive scattering length, a higher value of 5.5(2) is
measured showing deeper evaporative cooling. Essentially the same
$\eta$ of 5.3(2) is obtained at $B = 1290$\,G where scattering
takes place in the triplet-dominated regime with a very large
negative scattering length.

The observed behavior is in agreement with a Monte-Carlo
simulation taking into account the unitarity limitation of the
cross section \cite{mosk2002}. The results confirm the observed
evaporation time scales and the attained values of $\eta$. We also
find that collisions ejecting atoms out of the trap take place
with a mean energy of $\sim$\,350$\mu$K which corresponds to an
effective wave number of $k \approx (300\,a_0)^{-1}$.

\begin{figure}
\includegraphics[width=8cm]{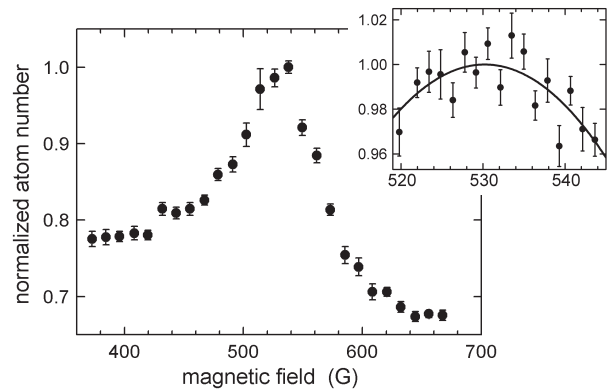}
\caption{\label{zoom}Measurements on plain evaporation in magnetic
fields close to the zero crossing of the scattering length under
same conditions as the in Fig.~\ref{plain} for a holding time of
3s. Here the number of remaining atoms is normalized to the
observed maximum value. The inset shows a series of measurements
in a very narrow range around the maximum at 530(3)\,G together
with a parabolic fit.}
\end{figure}

We measure the minimum-loss feature in a closer range of magnetic
fields to precisely determine the value of the magnetic field at
which the zero crossing of scattering length occurs. The main data
points in Fig.~\ref{zoom} are obtained with 500 individual
measurements at a holding time of 3\,s with the magnetic field
randomly varied between 30 values in an interval between 370\,G
and 670\,G; the data shown in the inset are obtained with 1000
measurements in the very narrow range between 520\,G and 544\,G.
The results allow us to determine the B-field for minimum
evaporative loss, and thus the zero-crossing of the scattering
length to $530(3)$\,G.

Forced evaporation measurements provide complementary data to
plain evaporation and allow us to rule out a significant role of
inelastic collisions. When the trap depth is ramped down, elastic
collisions reduce trap loss in contrast to increased loss at
constant trap depth. This can be understood by the spilling loss
of energetic particles \cite{Luiten1996a}: Without elastic
collisions the most energetic particles are spilled out of the
trap when its depth is reduced. With elastic collisions the
evaporative cooling effect decreases the temperature and thus
reduces the spilling loss.

\begin{figure}
\includegraphics[width=7cm]{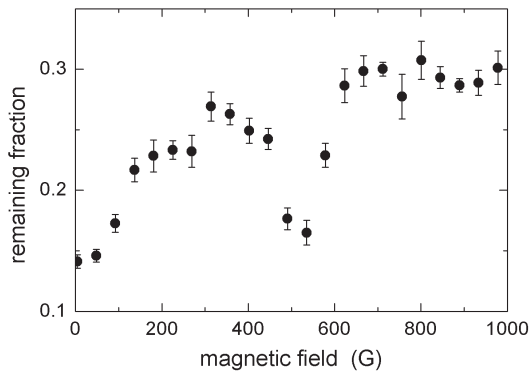}
\caption{\label{forced}Fraction of atoms remaining after forced
evaporation versus applied magnetic field. The trap depth is
ramped down exponentially to 20\% of the initial value in 10\,s.}
\end{figure}

In our forced evaporation measurements we reduce the trap depth in
10\,s to 20\% of its initial value in an exponential ramp and
measure the number of remaining atoms; the results are displayed
in Fig.~\ref{forced}. At reduced laser power of the REDT the
magnetic field has to be restricted to 1\,kG because of an
increasing sensitivity of the resonator lock to current-dependent
noise. A minimum number of atoms is now measured at 0\,G and
530\,G instead of the maximum observed with constant trap depth.
The largest number of atoms is observed in the high-field region
above 650\,G as expected for the large scattering cross section.

In Ref.~\cite{dieckmann2002} inelastic collisions in $^6$Li were
studied at a temperature of $\sim$$20\,\mu$K. A magnetic field
dependent loss was observed around 680 Gauss, with evidence of an
inelastic two-body process. To measure the loss rate under our
conditions of higher temperature and much lower density we study
the long-time evolution of samples which are pre-cooled by forced
evaporation and then recompressed into a 750\,$\mu$K deep trap to
a temperature of $\sim$150\,$\mu$K and a density of $5 \times
10^{10}$ cm$^{-3}$. The magnetic field during the 60-s holding
time is alternated for successive data points between 680 G and
300 G, the latter value corresponding to a very low loss rate in
Ref.~\cite{dieckmann2002}. We find no significant difference in
the number of remaining atoms, which leads to a clear upper bound
to the two-body rate constant of $1 \times 10^{-12}$ cm$^3/$s.
This upper bound is lower than the value of $2 \times 10^{-12}$
cm$^3/$s of Ref.~\cite{dieckmann2002}, whereas for a process
involving higher partial waves one would expect the rate to
increase with temperature. From our upper bound for inelastic loss
and the measured values for $\eta$ we infer that the ratio of
good-to-bad collisions at very large positive and negative
scattering length clearly exceeds 1000. Thus excellent conditions
for evaporative cooling towards degeneracy can be expected near
the Feshbach resonance.

In conclusion, our measurements confirm the predicted magnetic
tunability of the $s$-wave scattering length in a spin-mixture of
fermionic lithium atoms in the whole magnetic field range of
experimental interest. The observed zero crossing of the
scattering length at 530(3)\,G together with the large cross
section observed for higher fields
provides clear evidence of the predicted Feshbach resonance.
Moreover it enables ´more precise calculations of the $^6$Li
scattering properties. The resonance itself is masked by
unitarity-limited scattering and requires much deeper evaporative
cooling for a direct observation. The fact that we do not see any
significant effect of inelastic loss highlights that the extremely
large positive and negative scattering lengths attainable with
fermionic lithium offer intriguing new possibilities for
experiments on interacting Fermi gases.

{\it Note added:} Shortly before submission of the present
manuscript we learned about measurements of the group of J.E.\
Thomas on the zero crossing of the scattering length which agree
with our data.

\begin{acknowledgments}
We thank R.G.\ Hulet and H.\ Stoof for very useful discussions and
V.\ Venturi for valuable input. Support by the Austrian Science
Fund (FWF) within project P15115 and SFB\,15 (project part 15) and
by the Institut f\"ur Quanteninformation GesmbH is gratefully
acknowledged.
\end{acknowledgments}


\end{document}